\documentclass[11pt,prd,superscriptaddress,nofootinbib]{revtex4}
\usepackage[usenames,dvipsnames]{xcolor}

\usepackage{amssymb,amsmath,amsfonts}
\usepackage{graphicx}

\begin{document}

\title{Nucleon's axial-vector form factor in the hard-wall AdS/QCD model}

\author{Shahin Mamedov}
\email{sh.mamedov62@gmail.com}
\affiliation{Physics Dept, Gazi University, Teknikokullar, Ankara, Turkey}
\affiliation{Institute for Physical Problems, Baku State University, Z.Khalilov 23, Baku, AZ-1148, Azerbaijan}
\affiliation{Institute of Physics, Azerbaijan National Academy of Sciences, H.Javid ave. 33, AZ-1143, Baku, Azerbaijan}
\author{Berin Belma Sirvanli}
\email{bbelma@gazi.edu.tr}
\affiliation{Physics Dept, Gazi University, Teknikokullar, Ankara, Turkey}
\author{Ibrahim Atayev}
\email{atayevibrahim@gmail.com}
\affiliation{Institute of Physics, Azerbaijan National Academy of Sciences, H.Javid ave. 33, AZ-1143, Baku, Azerbaijan}
\author{Narmin Huseynova}
\email{nerminh236@gmail.com}
\affiliation{Institute for Physical Problems, Baku State University, Z.Khalilov 23, Baku, AZ-1148, Azerbaijan}
\affiliation{Theoretical Physics Department, Physics Faculty, Baku State University, Z.Khalilov 23, Baku, AZ-1148, Azerbaijan}

\maketitle

\centerline{\bf Abstract} \vskip 4mm
The axial-vector form factor of the nucleons is considered in the framework of hard-wall model of holographic QCD. A new interaction term between the bulk gauge and matter fields was included into the interaction Lagrangian. We obtain the axial-vector form factor of nucleons in the boundary QCD from the bulk action using AdS/CFT correspondence. The momentum square dependence of the axial-vector form factor is analysed numerically.
\vspace{1cm}
\section{Introduction}
The weak processes such as $\beta$ decay  $n \rightarrow p + e^{-}+ \tilde{\nu}_{e}$, $\mu$ capture $\mu^{-}+p=\nu_{\mu}+n$ and so on proceed via vector and axial-vector current of the nucleons. Form factors, which describe interaction vertex in these processes, are called the axial-vector form factors of the nucleons and were studied in the field theory framework. After the appearance of the QCD models based  on AdS/CFT correspondence \cite{1,2,3,4}, which were called holographic QCD or AdS/QCD models \cite{5,6,7,8,9,10,11}, these form factors were considered in the framework of these models as well. The AdS/QCD models such as hard-wall and soft-wall models were applied to the calculation of mass spectra, couplings and decay constants, form factors etc, which can be measured in the experiment \cite{12,13,14,15,16,17,18,19,20,21}. These models proved to be useful for the strong interaction studies in dense nuclear medium \cite {22,23,24,25,26} and at finite temperatures as well \cite{27,28,29,30,31,32,33}. Different form factors of mesons and  baryons were studied within the AdS/QCD models in \cite{12,13,14,15,16,17,18,34}.  The axial vector form factor of nucleons was considered in the framework of the soft-wall model of AdS/QCD in \cite{16}. Here we aim to consider this form factor in the hard-wall model framework.
In the second section we briefly present the QCD definition of the axial-vector form factor, which describes the nucleon-axial-vector field interaction vertex. In the third section we present basic features of the holographic hard-wall model: we introduce the scalar, axial-vector and fermion fields in the bulk of AdS space and write down their profile functions. In the fourth section  we establish the interaction Lagrangian between the bulk fields, which contributes to the axial-vector form factor and using the AdS/CFT correspondence we obtain an integral expression for this form factor. In the fifth section we fix parameters and perform a numerical analysis for this form factor. In the last section we summarise our result.
\section{Axial-vector current in QCD}
Isovector axial-vector current of nucleons in QCD is given by following definition \cite{35}:
\begin{equation}
j^{\mu ,a}\left( x\right) =\overline{\psi }\left( x\right) \gamma ^{\mu
}\gamma^{5}\frac{\tau ^{a}}{2}\psi \left( x\right).  \label{1}
\end{equation}%
Here  $\psi $ denotes the doublet of $u$ and $d$ quarks $\psi =\left(
\begin{array}{c}
u \\
d
\end{array}
\right)$ and $\tau^a$ are the Pauli matrices describing isospin.
This current is a partially conserved one, i.e.:
\begin{equation}
\partial _{\mu }j^{\mu ,a}\left( x\right) =i\overline{\psi }\left( x\right)
\gamma^{5}\left\{ \frac{\tau^{a}}{2},\mu \right\} \psi \left( x\right).\label{2}
\end{equation}
where $\mu $ denotes the mass matrix of the up and down quarks: $\mu =diag\ \left( m_{u},m_{d}\right) $. In the exact isospin symmetry case $\left( m_{u}=m_{d}=m\right) $ the $\mu$ matrix is proportional to the identity matrix. A matrix element of the isovector
axial-vector current between one-nucleon states is expressed in terms of form factors as follows \cite {35}:
\begin{equation}
\left\langle N\left( p^{\prime }\right) \left| j^{\mu ,a}\left(
0\right) \right| N\left( p\right) \right\rangle =\overline{u}\left(
p^{\prime }\right) \left[ \gamma ^{\mu }\gamma^{5}G_{A}\left( q^{2}\right) +
\frac{q^{\mu }}{2m_{N}}\gamma^{5}G_{P}\left( q^{2}\right) \right] \frac{\tau ^{a}}{2}u\left( p\right) .  \label{3}
\end{equation}%
Here $q_{\mu }=p_{\mu }^{\prime }-p_{\mu }$ is the total momentum in the interaction vertex  and $m_{N}$ is the nucleon mass, $G_{A}\left( q^{2}\right) $ and $G_{P}\left(q^{2}\right) $ are called the axial-vector and induced pseudoscalar form factors respectively. Due to Hermiticity of the $j^{\mu ,a}$ current the form factors $\ G_{A}\left( q^{2}\right) $ and $G_{P}\left(q^{2}\right) $ are real functions of $q^{2}$ in the $q^{2}\leq 0$ domain. In present work we study the $\ G_{A}\left( q^{2}\right) $ form factor in the hard-wall model of AdS/QCD.

\section{The AdS/QCD model}

Let us briefly describe the hard-wall model of the AdS/QCD introduced in~\cite{5,6,7,8,9,10} and developed in \cite{12,13,14,15,18}. Background geometry for the dual gravity theory in this model is given by the 5-dimensional (5D) anti de-Sitter space (AdS$_{5}$) with the metric chosen in Poincare coordinates:
\begin{equation}
ds^{2}=\frac{1}{z^{2}}\left( \eta _{\mu \nu }dx^{\mu }dx^{\nu
}-dz^{2}\right).  \label{4}
\end{equation}
The fifth coordinate $z$ extends from $0$ to $\infty $,
which are called the ultraviolet (UV) and infrared
(IR) boundaries of the AdS space, respectively. $\eta _{\mu \nu }$ is the metric of the 4D
flat Minkovski space $\left( \eta _{\mu \nu }=diag\left( 1,-1,-1,-1\right)
,\ \mu,\nu =0,1,2,3\right) $. The hard-wall model of AdS/QCD is based on the two cut-offs  of the $z$ coordinate, first one of which is the cut-off at the bottom of the AdS space by small $\varepsilon $ $(\varepsilon\rightarrow0)$ in order to avoid singularity of the metric (\ref{4}) at the $z \rightarrow 0$ limit and another one is the cut-off at the top of this space, i.e. at some $z_{m}$ value of this coordinate, which is considered as the free parameter of the model. The latter cut-off breaks conformal symmetry in the $\varepsilon\leq z \leq z_{m}$ slice of the AdS$_{5}$ space and guarantees the confinement property of the strongly interacted particles in the boundary QCD. The $z_m$ parameter is usually taken as $z_{m}=1/\Lambda_{QCD}$ ($\Lambda_{QCD}$ corresponds to the confinement scale of QCD) or it is fixed in such a way as to get exact mass spectra or other quantities (couplings, decay constants, form factors, etc) of the particles included into the problem. Thus, in the hard-wall model presented here the $z$ coordinate varies in the limited interval $\varepsilon \leq z\leq z_{m}$.

In order to obtain the axial vector form factor of nucleons in the boundary QCD we need to introduce fields in the bulk, whose boundary values correspond to the corresponding particles in the boundary QCD. In the AdS/QCD models  as a flavor symmetry group in the bulk of AdS space (\ref{4})  $SU(2)_L\times SU(2)_R$ group is chosen and the left and right chiral gauge fields $A_L$ and $A_R$ are introduced, which transform under the fundamental representations of this group $\left(1/2,0\right)$ and $\left(0,1/2\right)$ respectively. The bulk axial-vector field $A^{M}\left(x,z\right)=\frac{1}{\sqrt{2}}\left(A_L^M-A_R^M\right)$ is composed from these fields. Another field introduced in the bulk theory is the scalar $X$ field, which is transforming under the bifundamental $(1/2,1/2)$ representation of symmetry group and performs spontaneous symmetry breaking $SU(2)_L\times SU(2)_R\rightarrow SU(2)_V$. According to the AdS/CFT correspondence $SU(2)_V$ symmetry group of the bulk theory corresponds to the isospin symmetry of the boundary theory. In addition we introduce the spinor fields $\Psi_1$ and $\Psi_2$, both having left $\Psi_L$ and right $\Psi_R$ components in the bulk. Nucleons in the boundary QCD are described by means of boundary values of the bulk $\Psi$  fields. From the fields $A_M$, $X$ and $\Psi$ one can construct different kind interactions in the bulk and we shall choose among them only terms that will contribute to the axial-vector form factor of the nucleon. The axial-vector gauge field $A_{\mu}^{a}\left( x,z\right) $ in the bulk theory and the
 axial-vector current of spin-1/2 field $J_{\mu }^{a}\left( x\right) =
\overline{\psi}\left( x\right)\gamma^5 \gamma _{\mu }\frac{\tau^{a}}{2}\psi\left( x\right) $ in the boundary theory are the holographic duals.
Let us briefly present here bulk-to-boundary propagators (also are called profile functions) for the bulk fields, which are the solutions to the five dimensional equations of motion in the free field limit.

\subsection{Axial-vector field in AdS space}

The action for the bulk gauge fields $A_L$ and $A_R$ is written in the following form~\cite{13}:
\begin{equation}
S=-\frac{1}{2g_{5}^{2}}\int d^{5}x\sqrt{G}Tr\left( \mathcal{F}_{L}^{2}+
\mathcal{F}_{R}^{2}\right) ,  \label{5}
\end{equation}
where the field strength tensors $\mathcal{F}_{L,R}$ are defined as $\mathcal{F}_{MN}=\partial _{M}A_{N}-\partial _{N}A_{M}-i\left[ A_{M}, A_{N}\right] $ and  $G$ is the determinant of metric tensor $G_{MN}$. 5D gauge coupling constant $g_{5}$ is related to the number of colours $N_{c}$ $\left(g_{5}^2=\frac{12\pi^{2}}{N_{c}}\right)$  \cite{5} and for the $SU(2)_V$ boundary gauge symmetry group is $2\pi $. The action (\ref{5}) contains vector field and axial vector field parts and for our aim we need only in the last one. Axial gauge $A_5=0$ is chosen as a gauge condition on $A_5$ component. The equation of motion does not contain mixed derivatives and using separation anzats
\begin{equation}
\widetilde{A}_{\mu }\left( q,z\right) =A_{\mu }\left(
q\right) \frac{A\left( q,z\right) }{A\left( q,\varepsilon \right) }
\label{6}
\end{equation}
from the action (\ref{5}) we obtain the following equation of motion for the $A\left( q,z\right)$ Fourier component:
\begin{equation}
z\partial _{z}\left( \frac{1}{z}\partial _{z}A\left( q,z\right) \right)
+q^{2}A\left( q,z\right) =0.
\label{7}
\end{equation}
The ultraviolet (UV) and infrared (IR) boundary conditions on this solution are $A\left( q,\varepsilon\right) =1$ and $\partial_{z}A\left( q,z=z_{m}\right) =0$ respectively. Solution to this equation is expressed via first kind Bessel functions $J_{m}$ and $
Y_{m}$~\cite{13}:
\begin{equation}
A\left( q,z\right) =\frac{\pi }{2}qz\left[ \frac{Y_{0}\left( qz_{m}\right) }{
J_{0}\left( qz_{m}\right) }J_{1}\left( qz\right)-Y_{1}\left( qz\right)
\right],
\label{8}
\end{equation}
where the integration constants were found by using UV and IR boundary conditions.
It should be noticed, bulk to boundary propagator (\ref{8}) presented here was obtained for the free gauge field, while the considered here axial-vector field $A_M$ interacts with other fields included into the model. We shall neglect the back reaction of these fields and use the free bulk to boundary propagator (\ref{8}) for the form factor calculation.

\subsection{Fermion fields in the bulk and Nucleons on the boundary}

Nucleons in QCD  have the left- and right-handed components and in order to describe them in the boundary theory of the AdS/CFT correspondence two spinor fields $\Psi _{1}$ and $\Psi_{2}$ need to be introduced in the bulk theory , which have left and right handed components~\cite{12,14,15,16,17,18}. One of these fields $\left(\Psi_{1}\right)$ is necessary for the description of the left-handed component of the nucleon doublet in the boundary theory and the second one $\left(\Psi_{2}\right)$ gives the right-handed component of this doublet. Extra components (two of total four) of the bulk spinors are eliminated by the IR boundary conditions and this yields a mass spectrum of excited states of the nucleons. For clarity let us briefly present some formulas from this AdS/CFT correspondence constructed in~\cite{15}.

The action of bulk Dirac fields $\Psi_{1,2}$ has the form below:
\begin{equation}
S=\int d^{5}x\sqrt{G}\left[ i\overline{\Psi}_1 e^M_{A}\Gamma^{A}D_{M}\Psi_1-m_{5}\overline{\Psi }_{1}\Psi_{1}+\left(\Psi_1\leftrightarrow \Psi_2, \quad m_5\leftrightarrow-m_5 \right)\right].  \label{9}
\end{equation}
The covariant derivative $D_{B}$ is defined as follows:
\begin{equation}
D_{B}=\partial _{B}-\frac{i}{4}\omega _{B}^{MN}\Sigma _{MN}-i\left( A_{L}^{a}\right) _{B}T^{a},\label{10}
\end{equation}
where $\Sigma _{MN}=\frac{1}{i}\Gamma _{MN}$ and $\Gamma_{MN}=\frac{1}{2}\left[\Gamma_{M},\Gamma_{N}\right]$. For the metric (\ref{1})
the vielbein $e_{M}^{A}$ is chosen as $e_{M}^{A}=\frac{1}{z}\eta _{M}^{A}$
and the spin connection $\omega _{B}^{MN}$ has non-zero components
$\omega _{\mu }^{5A}=-\omega _{\mu }^{A5}=\frac{1}{z}\delta _{\mu }^{A}$ $
\left( \mu =0,1,2,3\right) $.

The equation of motion obtained from the action (\ref{9}) is the 5D
Dirac equations in the AdS space (\ref{1}):
\begin{equation}
\left(i\Gamma ^{A}D_{A} \mp m_{5}\right)\Psi_{1,2}=0,  \label{11}
\end{equation}
The boundary term which arise on obtaining the equation of motion is the following one:
\begin{equation}
\left(\delta \overline{\Psi}_1e^5_A\Gamma^A \Psi_1\right)|_{\varepsilon}^{z_m}=0.
\label{12}
\end{equation}
 5D $\Gamma $ matrices are usually chosen in the chirality basis~\cite{15},
\begin{equation}
\Gamma ^{5}=-i\gamma ^{5}=\left( \begin{array}{cc}
-i & 0 \\
0 & i
\end{array}
\right) ,\quad \Gamma ^{0}=\left( \begin{array}{cc}
0 & -1 \\
-1 & 0
\end{array}
\right) ,\quad \Gamma ^{i}=\left( \begin{array}{cc}
0 & \sigma ^{i} \\
-\sigma ^{i} & 0
\end{array}
\right) ,\quad \left( i=1,2,3\right) .  \label{13}
\end{equation}
A condition for elimination of the extra $\Psi_z$ degrees of freedom is chosen as  $\Psi_{z}=0$ ~\cite{15,18}. The left- and right-handed components of the $\Psi_i$ fields have properties $\gamma ^{5}\Psi _{L}=\Psi_{L}$ and $\gamma ^{5}\Psi _{R}=-\Psi _{R}$ and Fourier transformation for them are expressed in terms of 4D spinor field $\psi\left(p\right)$
\begin{equation}
\Psi _{L,R}\left( x,z\right) =\int d^{4}p\ e^{-ip\cdot
x}F_{L,R}\left( p,z\right) \psi _{L,R}\left( p\right)  \label{14}
\end{equation}
and the 4D spinor $\psi_{1}\left( p\right) $ obeys the 4D Dirac equation
\begin{equation}
\not{\!}{p}\psi_{1}\left( p\right) =\left\vert p\right\vert \psi_{1}\left( p\right) .  \label{15}
\end{equation}
Here, $\left\vert p\right\vert =\sqrt{p^{2}}$ for a time-like four-momentum $
p $. Then the 5D Dirac equation (\ref{11}) will lead to equations over the fifth coordinate $z$ for $F_{L,R}$ amplitudes:
\begin{equation}
\left( \partial _{z}^{2}-\frac{4}{z}\partial _{z}+\frac{6\pm m_{5}-m_{5}^{2}
}{z^{2}}\right) F_{L,R}=-p^{2}F_{L,R}.  \label{16}
\end{equation}

For consistency with the chirality of the dual boundary operator the sign of $m_5$ was chosen\footnote{See \cite{14,15,18} for more details.} positive $\left(m_5>0\right)$ for  $\Psi_{1}$ and negative $\left(m_5<0\right)$ for $\Psi _{2}$. The UV boundary conditions which are imposed on $F_{1L}$ and on $F_{2R}$ are
\begin{equation}
F_{1L}\left( p,\varepsilon \right) =0, \qquad F_{2R}\left( p,\varepsilon \right) =0,
\label{17}
\end{equation}
when $\varepsilon \rightarrow0$. Normalized solutions of equation (\ref{16}) for non-zero modes $(|p|\neq 0)$ are found using the UV boundary conditions and are expressed in terms of Bessel functions of first kind
\begin{equation}
F_{L,R}=C_{1,2}z^{5/2}J_{m_{5}\mp \frac{1}{2}}\left( \left\vert p\right\vert
z\right) ,  \label{18}
\end{equation}
where $C_{1,2}$ are normalization constants. The value of $m_{5}$ can be found from the relation $\left\vert m_{5}\right\vert =\Delta _{1/2}-2$, where scaling dimension $\Delta _{1/2}$ for the composite baryon operator is $\Delta _{1/2}=9/2$~\cite{18} and $\left\vert m_{5}\right\vert =5/2$. Consequently, for the $\Psi_{1,2}$ spinors the $ m_{5}$ "mass" have the values $m_{5}=\pm 5/2$, respectively. Thus, the $F_{1L,R}$ and $F_{2L,R}$ profile functions are given by
\begin{eqnarray} \label{19}
F_{1L}=C_{1}z^{5/2}J_{2}\left( \left\vert p\right\vert
z\right) , \qquad F_{1R}=C_{2}z^{5/2}J_{3}\left( \left\vert p\right\vert
z\right); \\ \nonumber
F_{2L}=-C_{2}z^{5/2}J_{3}\left( \left\vert p\right\vert
z\right) , \qquad F_{2R}=C_{1}z^{5/2}J_{2}\left( \left\vert p\right\vert
z\right).
\end{eqnarray}
As is seen from (\ref{19}) $F_{1L,R}$ and $F_{2L,R}$ are related one with another\footnote{This relation takes place only for the nucleons in ground states.}:
\begin{equation}
F_{1L}=F_{2R} , \qquad F_{1R}=-F_{2L}.
\label{20}
\end{equation}
For obtaining only a left-handed component of the nucleon from $\Psi_1$ the right-handed component of this spinor is eliminated by the boundary condition at $z=z_{m}$:
\begin{equation}
\Psi _{1R}\left( x,z_{m}\right) =0.  \label{21}
\end{equation}
This condition gives the Kaluza-Klein mass spectrum $M_{n}$ of excited states, which is expressed in terms of zeros $\alpha _{n}^{(3)}$ of the Bessel function $J_{3}$:
\begin{equation}
M_{n}=\frac{\alpha _{n}^{(3)}}{z_{m}}.
\label{22}
\end{equation}
The quantum number $n$ corresponds to the excitation number of a nucleon in the dual boundary theory.

Similarly, the right-handed component of the nucleon can be obtained from the $\Psi_2$ spinor imposing the boundary condition on the left-handed component of this field, which again leads to mass spectrum (\ref{22}).

The normalization constants $C_{1,2}$ in (\ref{19}) are equal and for the $n$-th excited state were found as follows ~\cite{14}:
\begin{equation}
|C_{1,2}^n|=C^n=\frac{\sqrt{2}}{z_{m}J_{2}\left( M_{n}z_{m}\right) }.
\label{23}
\end{equation}

In the AdS/CFT correspondence for spinor field there is another way for eliminating the boundary terms (\ref{12}), which consists in introducing additional terms to the action (\ref{9}) \cite{42,43,44}. These terms are boundary terms and are equal to the boundary terms (\ref{12}), but have opposite sign \cite{45}. This trick is used in the soft-wall model case as well \cite{12}, where the $z$ direction of AdS space extends to infinity. The trick with introducing  of  boundary terms can be useful in solving problems in the hard-wall model as well.

As is seen from (\ref{22}) the mass spectrum is determined by the value of $z_m$, which is free  parameter of the model. Spectra of all fields, which were included into the hard-wall model, are determined  by this parameter. We shall see later the  axial-vector form factor's expression depends on this parameter, as well. There is a difficulty to fit the values of all spectra, coupling constants, form factors etc. to the experimental ones by fixing unique free parameter of the model. But this difficulty in hard-wall model can be easily overcame by introducing extra boundary terms. To this end, we should introduce extra terms at infrared boundary at the same time with applying the boundary conditions at this boundary\footnote{Such a trick was used in \cite{26}, where different isocomponents of the same field in the isospin background had different quantization formulas of spectra because of isospin interaction with the background. Quantization formulas for the spectra of all isocomponents, which were obtained from the boundary condition at $z=z_{IR}$, were reduced to the same one and the values of the infrared boundaries for the different isocomponents were reduced to the same $z_{IR}$ by introducing extra infrared boundary terms.}. Remind, that additional boundary terms do not depend on $z$ and consequently, do not change the equations of motions. If we want to shift the mass spectrum (\ref{22}) from the value $z_m$ to some value $z_1$, in which the new spectrum will be coincident with the experimental data, we can add boundary term written below\footnote{It should be noted that, in \cite{43} a choice of $\Gamma$ matrices is different than here and the $\Gamma^5$ in that work has a form $
\Gamma ^{5}=-i\gamma ^{5}=\left( \begin{array}{cc}
0 & i \\
i & 0
\end{array}\right)$. This is a reason of distinction between the forms of the extra boundary term here and in \cite{45}.} to the action (\ref{9}):
\begin{eqnarray}
 &S_{bdy}=\int d^{4}x\sqrt{G} \overline{\Psi}_1\left(z\right)e^5_A\Gamma^A\left[ \Psi_1\left(z_1\right)-\Psi_1\left(z\right)\right]|_{z=z_m} \nonumber \newline& \\
&=-i\int d^{4}x\frac{1}{z^4}\left\{ \overline{\Psi}_{1L}\left(z\right)\left[ \Psi_{1L}\left(z_1\right)-\Psi_{1L}\left(z\right)\right]-\overline{\Psi}_{1R}\left(z\right)\left[ \Psi_{1R}\left(z_1\right)-\Psi_{1R}\left(z\right)\right]\right\}|_{z=z_m}.\label{24}
\end{eqnarray}
It is obvious, that the variation $\delta S_{bdy}$ will be summed with the boundary term (\ref{12}) and the $\Psi_1\left(z\right)$ terms in the square brackets above will cancel with it. Imposing a boundary condition on the remaining boundary term we get following boundary conditions:
\begin{equation}
\left(\delta \overline{\Psi}_1\left(z\right)e^5_A\Gamma^A \Psi_1\left(z_1\right)\right)|_{\varepsilon}^{z_m}=0.
\label{25}
\end{equation}
This leads to boundary condition on $\Psi_1$ at $z=z_1$
\begin{equation}
\Psi _{1R}\left( x,z_{1}\right) =0,  \label{26}
\end{equation}
and the mass spectrum (\ref{22}) now will be determined by $z_1$. In such a way we can shift mass spectrum  not shifting the infrared boundary of the model. The condition (\ref{26}) will give a desirable spectrum for the spinor field in the case when a right form factor shape will be obtained at $z=z_m$ while right mass spectrum is obtained at $z=z_1$.

\subsection{Scalar field responsible for the chiral symmetry breaking}
Action for the scalar $X$ field is the usual one for the scalar field in the five dimensional background geometry (\ref{4}):
\begin{equation}
S=-\int d^{5}x\sqrt{G} Tr \left[\left| D X \right|^2 + {3} \left| X \right|^2 \right],
\label{27}
\end{equation}
The covariant derivative $D_M$ includes interaction of this field with the gauge fields $A_L$ and $A_R$:
\begin{equation}
D_M X = \partial_M X - i \left(A_L\right)_M X + i X \left(A_R\right)_M.
\label{28}
\end{equation}
Its interaction with the spinor fields will be written in separate terms in the interaction Lagrangian. The $X$ field is written in the form: $X=v\left(z\right)\exp\left[i\sqrt{2}\pi^aT^a\right] $, where the $\pi^a$ field in the dual QCD describe pions.
In the free field limit the solution of the equation of motion obtained from the action (\ref{27}) for $v\left(z\right)$ has a form \cite{14}:
\begin{equation}
v\left(z\right)=\frac{1}{2}m_{q}z+\frac{1}{2}\sigma z^3,\label{29}
\end{equation}
where  $m_q$ is the mass of bare light quarks and $\sigma$  is the value of chiral quark condensate. We shall use solution (\ref{29}) in our calculations of $G_A\left(q^2\right)$.
\section{Form factor from the hard-wall model}
\subsection{Interaction Lagrangian in the bulk and AdS/CFT correspondence for the axial-vector current}
There are different kinds of interactions between the $A_M$, $X$ and $\Psi_{1,2}$ fields in the bulk of the AdS space. The interaction Lagrangian of these fields contains  axial-vector current of the bulk spinor fields. The total interaction Lagrangian consist of the terms contributing to both vector and axial-vector currents. However, we are interested only in the interaction terms producing axial-vector $\overline{\psi }\left( x\right) \gamma ^{\mu}\gamma^{5}\frac{\tau ^{a}}{2}\psi \left( x\right)$ structure in the action. Namely such terms will contribute to the axial-vector form factor $G_{A}\left( q^{2}\right) $ in the AdS/CFT correspondence of the bulk theory and boundary QCD. Several such bulk interaction Lagrangian terms are presented in the literature \cite{14,16,18}. Let us list all these interactions between the bulk fields $A_{M}$, $X$ and $\Psi_{1,2} $ existing in the literature that contribute to the $G_{A}\left( q^{2}\right) $ form factor:

1)  a minimal coupling term
\begin{equation}
\textit{L}^{(1)}=\overline{\Psi }_{1}\Gamma ^{M}\left(A_{L}\right)_M\Psi _{1}-\overline{\Psi }_{2}\Gamma ^{M}\left(A_{R}\right)_M\Psi _{2}=\frac{1}{2}\left(\overline{\Psi }_{1}\Gamma ^{M}A_{M}\Psi _{1}-\overline{\Psi }_{2}\Gamma ^{M}A_{M}\Psi_{2}\right),\label{30}
\end{equation}
2)a magnetic gauge coupling term
\begin{eqnarray}
\textit{L}^{(2)}=ik_1\left\{\overline{\Psi }_{1}\Gamma^{MN}\left(F_L\right)_{MN}\Psi _{1}-\overline{\Psi}_{2}\Gamma^{MN} \left(F_R\right)_{MN}\Psi_{2}\right\}\nonumber\\ =\frac{i}{2}k_1\left\{\overline{\Psi }_{1}\Gamma^{MN}F_{MN}\Psi _{1}+\overline{\Psi }_{2}\Gamma^{MN}F_{MN}\Psi_{2}\right\},\label{31}
\end{eqnarray}
where $F_{MN}=\partial_MA_N-\partial_NA_M$ is the field stress tensor of the axial-vector field $A_M$.

3) a Yukawa type interaction term, which describe interaction of the bulk fermion fields with the scalar $X$ field
\begin{equation}
\textit{L}^{(3)}=g_Y\left(\overline{\Psi }_{1}X\Psi_{2}+\overline{\Psi }_{2}X^{\dagger }\Psi_{1}\right),  \label{32}
\end{equation}
This interaction flips a chirality of fermions $\Psi_{1,2}$ and in the dual theory refers to interaction of the nucleon with a chiral condensate.

4) An interaction term for the triple interaction of the fermion fields $\Psi_{1,2}$ with the scalar $X$ field and with the vector field $V$ or with the axial vector field $A$ was introduced in \cite{18}:
\begin{eqnarray}
\textit{L}^{(4)}=\frac{i}{2}k_{2}\left[\overline{\Psi }_{1}X \Gamma^{MN} \left( F_{R}\right) _{MN}\Psi _{2}+\overline{\Psi }_{2}X^{\dagger }\Gamma ^{MN}\left( F_{L}\right) _{MN}\Psi _{1}+h.c.\right] \nonumber \\
=-\frac{i}{2}k_{2}\overline{\Psi }_{1}X
\Gamma ^{MN} F _{MN}\Psi _{2}+\frac{i}{2}
k_{2}\overline{\Psi }_{2}X^{\dagger } \Gamma^{MN}F _{MN}\Psi _{1} + vector \ term
\label{33}
\end{eqnarray}
This term was interpreted as a magnetic type interaction of fermion fields with the vector and axial-vector fields, where the scalar field takes part as well \cite{21}.

5) Similar to $\textit{L}^{(4)}$ we compose a new interaction term, which is hermitian and is parity invariant:
\begin{eqnarray}
\textit{L}^{(5)}=\frac{g_Y}{2}\left[\overline{\Psi }_{1}X \Gamma
^{M}\left(A_{L}\right)_M\Psi_{2}-\overline{\Psi }_{2}X^{\dagger}
\Gamma ^{M}\left(A_{R}\right)_M\Psi_{1}+h.c.\right] \nonumber \\
=g_Y\left(\overline{\Psi }_{1}X\Gamma^{M}A_{M}\Psi_{2}+\overline{\Psi }_{2}X^{\dagger}\Gamma ^{M}A_{M}\Psi_{1}\right).
\label{34}
\end{eqnarray}
This term also describes a triple interaction of the bulk fields, but $\textit{L}^{(5)}$ is not magnetic type one in unlike $\textit{L}^{(4)}$. Since this interaction is the chirality changing one as a coupling constant in it was chosen the Yukawa coupling constant $g_Y$. Calculations show that $\textit{L}^{(5)}$  contributes only to the axial-vector form factor of nucleons.

6) Beside the terms presented above there is the minimal-type coupling term introduced in \cite{16} for calculation of the axial-vector form factor in the framework of soft-wall model of AdS/QCD:
\begin{equation}
 \textit{L}^{(6)}=\overline{\Psi }_{1}
\Gamma ^{M}\Gamma^{z}A_{M}\Psi_{1}+\overline{\Psi}_{2}
\Gamma ^{M}\Gamma^{z}A_{M}\Psi_{2}
\label{35}
\end{equation}
However, we omit this interaction since  these terms are not invariant with respect to  rotations in the $(z,x)$ plane in the five dimensional AdS space-time\footnote{In the five dimensional theory $\Gamma^z$  is the fifth component of the 5D $\Gamma^A$ vector and under rotations of the $\left(z,x\right)$ plane it is transformed as  $z$ component of this vector. So, it can not be considered as a scalar matrix in five dimensional theory as was  $\gamma^5$ in four dimensional theory. Consequently, the Lagrangian terms in (\ref{35}) do not remain  invariant under these rotations. So, this kind of interaction can not be considered as a physical one in the bulk theory. The interpretation "a minimal-type coupling, which is absent in four dimensions, but exists in five dimensions" concerning these terms is not consistent with the holographic duality, since this duality establishes a correspondence between the interactions in four and five dimensions both of which are physical.}.

Having an interaction Lagrangian in the bulk we can establish a holographic formula for the form factor. The AdS/CFT correspondence in our case matches the axial-vector current of  bulk fermions with the axial-vector current of nucleons in the boundary QCD and the bulk axial-vector field with the axial-vector meson in this boundary theory. So, having calculated the axial-vector current of bulk fermions we find the axial-vector current of nucleons. The bulk axial-vector  current can be found from the generating functional $Z$ which is defined as an exponent of the classical bulk action $S$:
\begin{equation}
Z_{AdS}=e^{iS_{int}}.
\label{36}
\end{equation}
Holography principle identifies the generating functional $Z_{AdS}$ of the bulk theory with the generating function $Z_{QCD}$ of the boundary QCD:
\begin{equation}
Z_{AdS}=Z_{QCD}.
\label{37}
\end{equation}
According to holographic equality (\ref{37}) the vacuum expectation value of the nucleon's axial-vector current in the boundary QCD theory can be found by taking variation from the gravity functional $Z_{AdS}$:
\begin{equation}
<J_{\mu}^a>^{QCD}=-i\frac{\delta Z_{AdS}}{\delta A_{\mu}^{a}}|_{A_{\mu}^{a}=0}
\label{38}
\end{equation}
We shall see that the formula (\ref{38}) will produce the axial-vector current $J_{\mu}^a(p^{\prime},p)=G_A\bar{u}(p^{\prime})\gamma^5\gamma_{\mu}\left(\tau^a/2\right)u(p)$, where $G_A$ denotes the integrals over the $z$ coordinate and is accepted as the axial-vector form factor of nucleons. In this holography a source for the $J_{\mu}^a$ current\footnote{In the exact isospin symmetry case the quark (and nucleon) masses are equal and $G_A $ is the same for all isospin states (for both nucleons). In the medium, when we deal with isospin asymmetry, it will be different for the different isospin states due to the mass splitting effect (\cite{24,25,26}).} will be axial-vector field $A_{\mu}^{a}$.

\subsection{The $G_A$ form factor}
The action $S_{int}$ is the 5D space integrals of the total interaction Lagrangian  $L_{int}=\sum_i L^{(i)}$. In Fourier components the 4D position integrals gives the usual $\delta$ function of energy-momentum conservation $q=p^{\prime}-p$. For a brevity of expressions calculated in the momentum space, we use the following notation for the $j^{5\mu}\left(p^{\prime},p \right)$ current in the action terms:
 \begin{equation}
j^{5\mu}\left(p^{\prime},p \right)=\overline{u}\left( p^{\prime }\right) \gamma
^{5}\gamma ^{\mu }\frac{\tau ^{a}}{2}u\left( p\right).
\label{39}
 \end{equation}
Let us list the action terms, where appears the $j^{5\mu}\left(p^{\prime},p \right)$ structure:
\begin{eqnarray}
& 1)\ S^{(1)}=\frac{1}{2}\int d^{4}x\ \int_{0}^{z_{m}}dz\ \sqrt{G}\left\{\overline{\Psi }_{1}\Gamma ^{\mu }A_{\mu }\Psi _{1}-\overline{\Psi }_{2}\Gamma ^{\mu
}A_{\mu }\Psi _{2}\right\}\nonumber \\
&=\frac{1}{2}\int d^{4}pd^{4}p^{\prime } j^{5\mu}\left(p^{\prime},p \right)A_{\mu }^{a}\left( q\right) \int_{0}^{z_{m}}dz\ \frac{1}{z^{4}}A\left( qz\right) \left[ \left\vert F_{1R}\left( mz\right) \right\vert
^{2}-\left\vert F_{1L}\left( mz\right) \right\vert ^{2}\right],
\label{40}
\end{eqnarray}
\begin{eqnarray}\label{41}
& 2)\ S^{(2)}=\frac{i}{4}k_1\int d^{4}x\ \int_{0}^{z_{m}}dz\ \sqrt{G} \left\{\overline{\Psi }_{1}\left[ \Gamma ^{5},\Gamma ^{\mu}\right]\partial_{5}A_{\mu}\Psi _{1}+ \overline{\Psi }_{2}\left[ \Gamma ^{5},\Gamma ^{\mu}\right]\partial_{5}A_{\mu}\Psi _{2}\right\}\nonumber \\
&=\frac{k_1}{2} \int d^{4}pd^{4}p^{\prime }j^{5\mu}\left(p^{\prime},p \right)A_{\mu }^{a}\left( q\right)\int_{0}^{z_{m}}dz\
\frac{1}{z^3}\left( \partial _{z}A\left( qz\right) \right) \left[ \left\vert
F_{1R}\left( mz\right) \right\vert ^{2}+\left\vert F_{1L}\left( mz\right)
\right\vert ^{2}\right],
\end{eqnarray}
\begin{eqnarray}
& 5)\ S^{(5)}=g_Y\int d^{4}x\ \int_{0}^{z_{m}}dz\ \sqrt{G}\left\{\overline{\Psi }_{1}X\Gamma ^{\mu }A_{\mu }\Psi _{2}+\overline{\Psi }_{2}X^{\dagger}\Gamma^{\mu}A_{\mu }\Psi _{1}\right\}\nonumber \\
&=g_Y\int d^{4}pd^{4}p^{\prime }j^{5\mu}\left(p^{\prime},p \right)A_{\mu }^{a}\left( q\right)\int_{0}^{z_{m}}dz \frac{1}{z^4}A\left( qz\right) 2 v\left(z\right) F_{1L}\left( mz\right) F_{1R}\left( mz\right).\label{42}
\end{eqnarray}
The AdS/CFT correspondence (\ref{37}) takes place on UV boundary of space-time and on this boundary the bulk fermions are considered on mass shell  $\left(|p|=|p^{\prime}|=m\right)$, which was taken into account in the profile functions $F_{1R}$ and $m$ is identified with the mass of nucleons due to this holography. It should be noted that $L^{(4)}$ may contribute to the axial-vector form factor when nucleons are in exited states and for them relations (\ref{20}) are not satisfied.

According to holographic formula (\ref{38}) the total action $S=S^{(1)}+S^{(2)}+S^{(5)}$ will produce the axial-vector form factor  $G_A$ of the  ground state nucleons.
Taking derivatives over $A_{\mu }^{a}\left( q \right) $ from the $S^{(i)}$ action terms we shall get contributions $G_A^{(i)}$ of these terms into the the axial-vector form factor $G_{A}\left(q^{2}\right)$:

\begin{equation}
1) \ G^{(1)}_{A}\left( q^{2}\right) =\frac{1}{2}\int_{0}^{z_{m}}dz\ \frac{1}{z^{4}}A\left(qz\right) \left[ \left\vert F_{1R}\left( mz\right) \right\vert^{2}-\left\vert F_{1L}\left( mz\right) \right\vert ^{2}\right],
\label{43}
\end{equation}
\begin{equation}
2) \ G^{(2)}_{A}\left( q^{2}\right) =\frac{k_1}{2}\int_{0}^{z_{m}}dz\frac{1}{z^3}\left(\partial_{z}A\left( qz\right) \right) \left[ \left\vert F_{1R}\left(mz\right) \right\vert ^{2}+\left\vert F_{1L}\left( mz\right) \right\vert ^{2}
\right],
\label{44}
\end{equation}
\begin{equation}
5) \ G^{(5)}_{A}=2 g_Y\int_{0}^{z_{m}}dz \
\frac{1}{z^4}A\left( qz\right)  v\left(z\right) F_{1L}\left( mz\right) F_{1R}\left( mz\right).
\label{45}
\end{equation}
In the next section we shall perform numerical analysis of the $G_A\left(Q^2\right)$  form factor  defined as the sum of $G_A^{(i)}$ terms:
\begin {equation}
G_A=G_A^{(1)}+G_A^{(2)}+G_A^{(5)}.
\label{46}
\end{equation}

\section{Numerical results}
We shall carry out a numerical study of the dependence of the $G_A$ form factor on $Q^2=-q^2$. In terms of $Q$ dependence the profile function  $A\left( qz\right)$ for  axial-vector field in (\ref{8}) gets the form below:
\begin{equation}
A\left(Qz\right)=\frac{\pi}{2}Qz\left[\frac{K_0\left(Qz_m\right)}{I_0\left(Qz_m\right)}
I_1\left(Qz\right)+K_1\left(Qz\right)\right],
\label{47}
\end{equation}
where $K_i\left(Qz_m\right)$ and $I_i\left(Qz_m\right)$ are the second kind Bessel functions. Using MATHEMATICA package we can  numerically calculate integrals $G_A^{(1)}\left(Q^2\right)$, $G_A^{(2)}\left(Q^2\right)$ and $G_A^{(5)}\left(Q^2\right)$  and present $Q^2$ dependence of $G_A\left(Q^2\right)$ form factor. For performing numerical integration for the value of light quark mass $m_q$ and for the value of condensate $\sigma$ were taken the values $m_q=0.00234 \ GeV$ and $\left(\sigma\right)^{1/3}=0.311 \ GeV$ \cite{14,18,46}. The constant $k_1=-0.98$ was taken from \cite{18}. This value was obtained from the fitting of couplings $g_{\pi NN}$  and $g_{\rho NN}$ obtained in the framework of hard-wall model with the experimental data. The value $g_Y=9.182$ was taken from \cite{46} and it was fixed to get correct nucleon mass within the hard-wall model having fixed parameters $m_q=0.00234 \ GeV$, $\left(\sigma\right)^{1/3}=0.311 \ GeV$ and $z_m=\left(0.330 \ GeV\right)^{-1} $. Here we have one more free parameter $z_1$ and we choose it in order to fit the mass $m$ of first Kaluza-Klein fermion mode $(n=1)$ to the ground state nucleon's mass $\left(m=0.94 \ GeV\right)$. This can be achieved by choice $z_1=\alpha^{(3)}_{1}/0.94=6.38/0.94 \approx \left(0.1473 \ GeV \right)^{-1}$. Thus, we choose the $m=0.94 \ GeV$ value for the $m$ parameter in the form factor expression. The value of free parameter $z_m$ is chosen  $z_m=\left(0.205 \ GeV\right)^{-1} $ \cite{14} and $z_m=\left(0.324 \ GeV\right)^{-1} $  \cite{18}. These values were chosen for obtaining proper mass spectrum of the mesons and nucleons within the hard-wall AdS/QCD model. In our numeric calculation we use the value of $z_m$ between these ones, i.e. $z_m=\left(0.286 \ GeV\right)^{-1}$ value. In Fig.1 we present the numerical result for $Q^2$ dependence of $G_A\left(Q^2\right)$ form factor (\ref{46}). As is seen from Fig.1 the hard-wall result for $G_A\left(Q^2\right)$ form factor dependence in $Q^2$ has a shape of $1/Q^n$ dependence.
\begin{figure}
\includegraphics[scale=0.30]{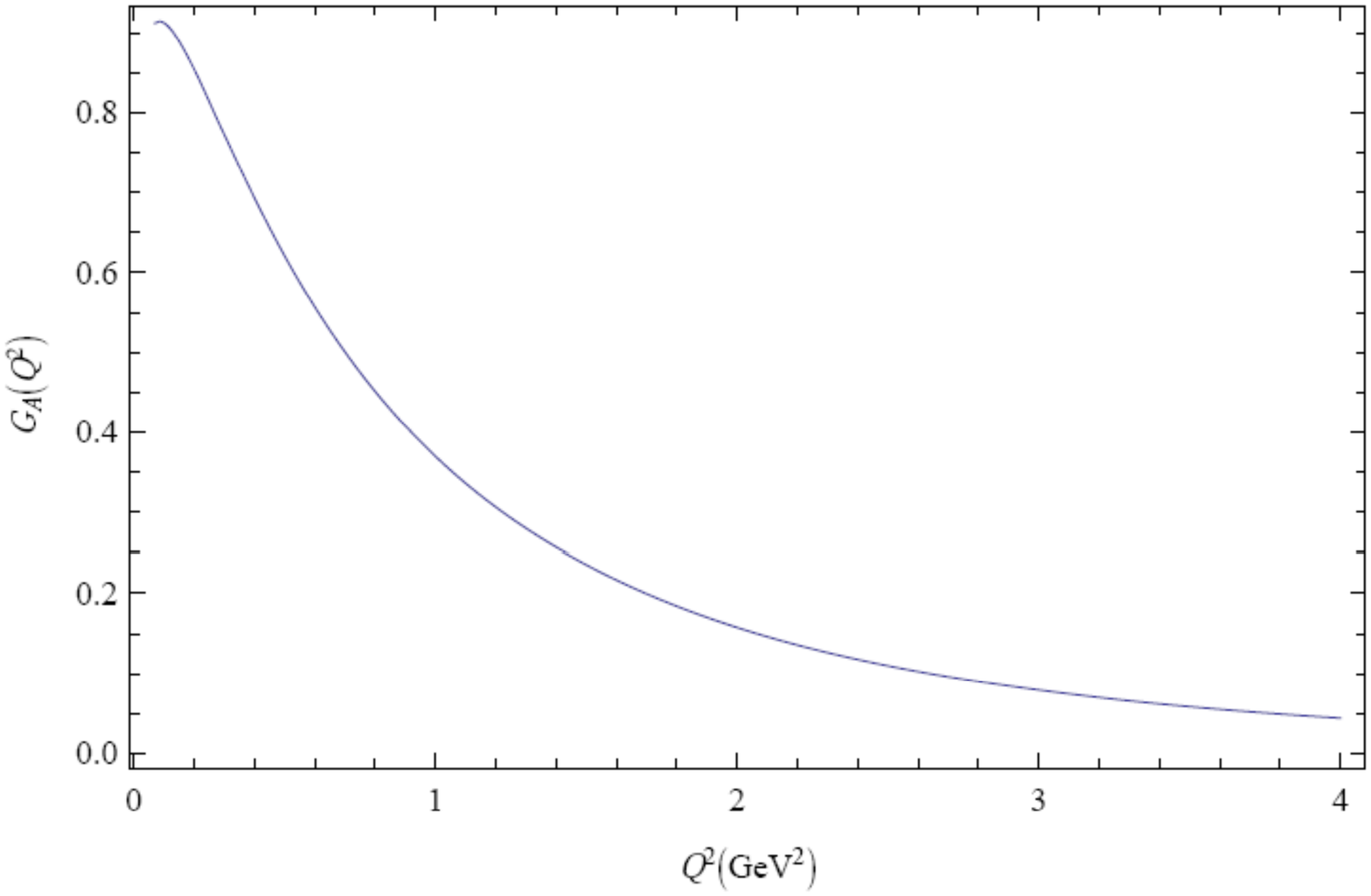}
\caption{Numerical result for $G_A\left(Q^2\right)$ form factor.  }
\label{fig:1}
\end{figure}
\begin{figure}
\includegraphics[scale=0.247]{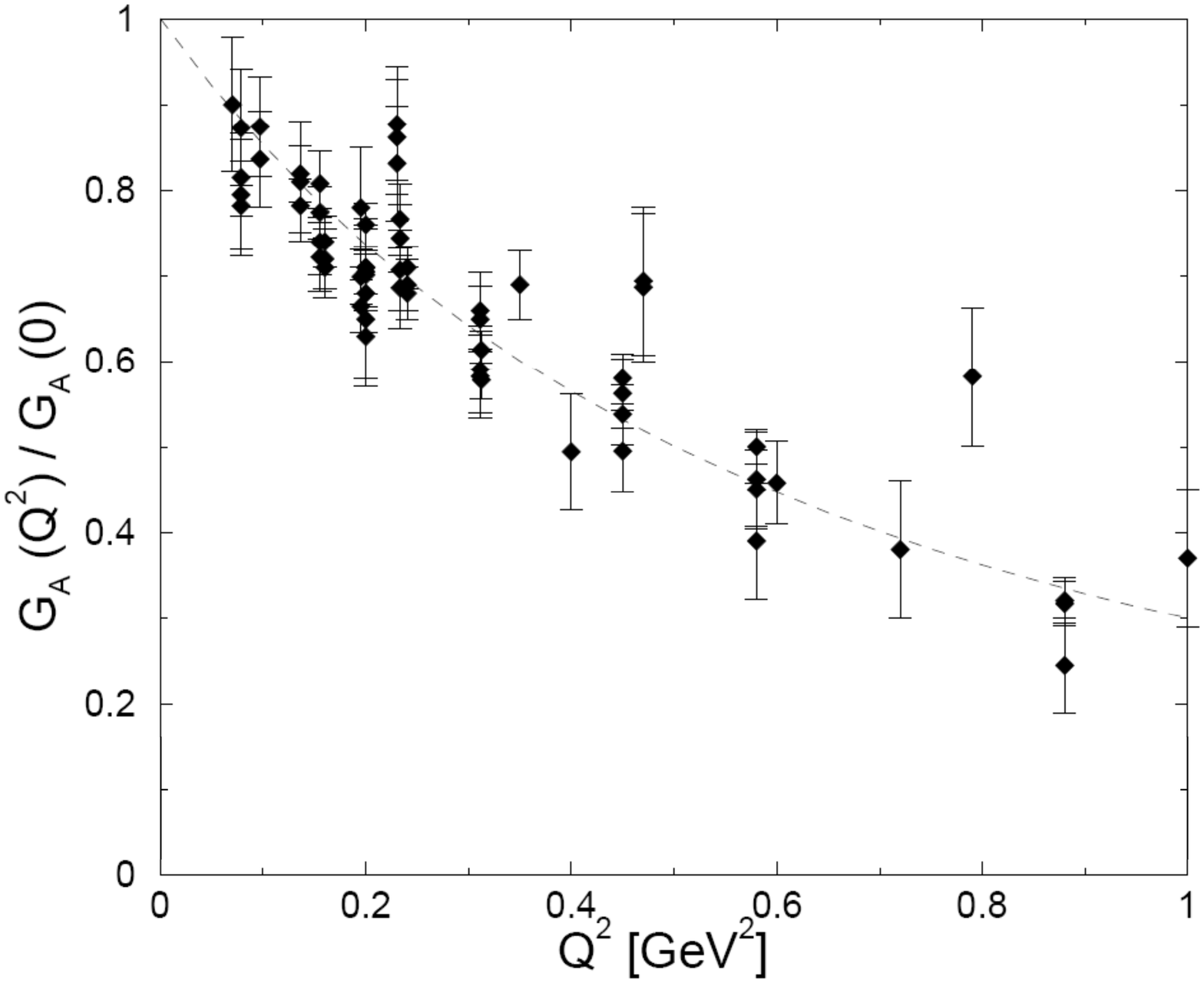}
\includegraphics[scale=0.25]{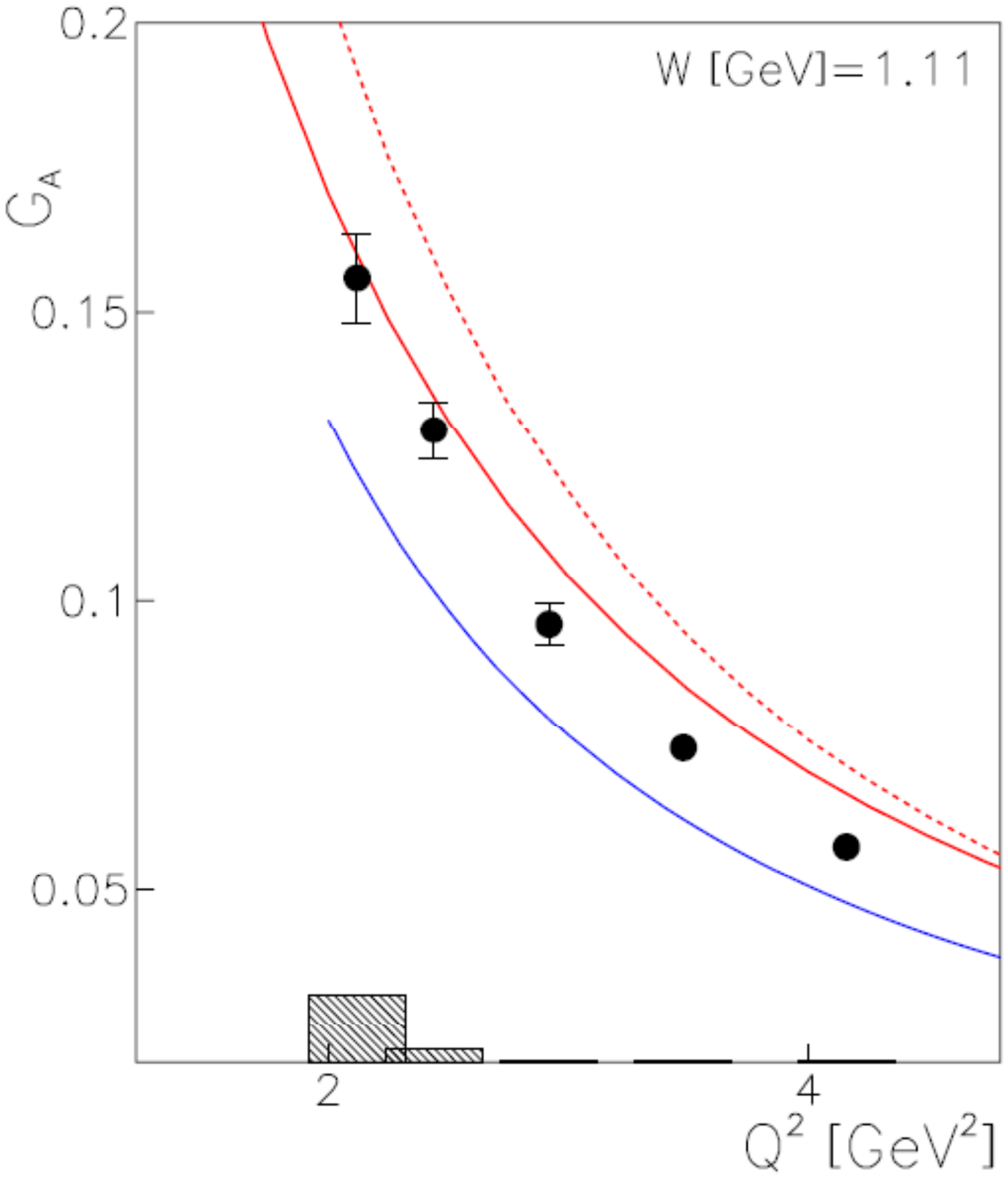}
\caption{{\it Left}: Experimental data for $G_A\left(Q^2\right)/G_A\left(0\right)$ form factor at small $Q^2$ taken from \cite{35}. {\it Right}: Experimental data for $G_A\left(Q^2\right)$ form factor at high $Q^2$ taken from \cite{40}. Blue solid line: MAID2007 for $G_A$, and red solid-dash lines: LCSR (red solid is the LCSR calculation using experimental electromagnetic form factors as input and red dash is pure LCSR) \cite{47}. }
\label{fig:2}
\end{figure}
 Such a dependence is typical for the nucleon form factors \cite{12,16,18,35,36,37,38,39,39,41}. Existing experimental data for $G_A$ form factor are in two different intervals of $Q^2$, in $\sim 0.075\leq Q^2 \left(GeV^2\right)\leq 1$ (\cite{35}) and in $2\leq Q^2 \left(GeV^2\right)\leq4$ (\cite{40}) intervals. These data are presented in Fig.2. The dependence in Fig.1 covers both these intervals. It is seen from the  Figs.1 and 2 the shapes of $G_A$ form factor dependence on $Q^2$  in these  figures are very close. For detailed analysis we make comparison with the experimental data and with other theoretical approaches in two tables. We presented in Tables 1 and 2 the values of $G_A$ form factor at several fixed values of $Q^2$ from the first and and from the second intervals, respectively\footnote{Note that the values of $G_A$ presented in the tables, except the hard-wall results, were taken from the graphs in the corresponding reference and are the approximate values.}. For a clarity we present in these tables the values of the $ G_A^{(1)} $, $ G_A^{(2)} $ and $ G_A^{(5)}$ terms as well. In  Table 1 we compare our hard-wall results with the experimental data and with the results of other approaches. In this table the results of the pion electroproduction data for  $G_A\left(Q^2\right)/G_A\left(0\right)$ were taken from \cite{35}, the results of Lattice Simulations calculations were taken from \cite{36,37} and the results obtained by Chiral Perturbation Theory application were taken from \cite{41}. In Table 2 the experimental data for $G_A$ were taken from \cite{40} and the results of the Light Cone Sum Rules calculations were taken from \cite{38}. It is seen from Table 1 the hard-wall results  are close to the experimental data from the pion electroproduction experiments and to the results of  Lattice Simulations and Chiral Perturbation theory in the interval of the small $Q^2$ values. From the Table 2 it is seen that the hard-wall results are close to the experimental data from CLASS collaboration and to the results of Light Cone Sum Rules in the interval of high $Q^2$ values. Numerical result for the value of $G_A\left(0\right)$ is ~0.65.
\begin{table}[!htb]
\begin{center}
\begin{tabular}{|c|c|c|c|c|c|c|c|}
\hline
 $Q^2 (GeV^2)$ & $G_A^{hard-wall}$ & $G_A^{exp.}\left(Q^2\right)/G_A\left(0\right)$ & $G_A^{lat.sm.}$ & $G_A^{ChPT}\left(Q^2\right)/G_A\left(0\right)$  & $ G_A^{(1)} $ & $ G_A^{(2)} $ & $ G_A^{(5)}$ \\
\hline
0.1 & 0.91 & 0.9 &0.9 & 0.83 &-0.28 &0.76 & 0.43 \\
\hline
0.2 & 0.85 &0.7 & 0.8& 0.72 & -0.23 & 0.48 & 0.60\\
\hline
0.3 &0.77 & 0.6& 0.7& 0.64 & -0.19 & 0.476 & 0.48 \\
\hline
0.4& 0.69 &0.5& 0.6 & 0.6 &-0.16 & 0.45 & 0.4 \\
\hline
0.6 & 0.55 &0.45 & 0.55 & -&-0.12 & 0.39 &0.28 \\
\hline
0.7 & 0.50 & 0.39 &0.5 &- &-0.10 & 0.36 & 0.24 \\
\hline
0.9 &0.4 &0.35 & 0.4 & -&-0.08 &0.31 & 0.18\\
\hline
\end{tabular}
\caption{The values  of the $G_A$ form factors at small values of $Q^2$
\label{tab:1}}
\end{center}
\end{table}

\begin{table}[!h]
\begin{center}
\begin{tabular}{|c|c|c|c|c|c|c|c|}
\hline
 $Q^2 (GeV^2)$ & $G_A^{hard-wall}$ & $G_A^{exp}$  & $G_A^{LCSR}$ ABO1 & $G_A^{LCSR} $ABO2 & $ G_A^{(1)} $ & $ G_A^{(2)} $ & $ G_A^{(5)}$ \\
\hline
2 & 0.158 & 0.16 & 0.111 & 0.128 & -0.03 & 0.17 & 0.05 \\
\hline
2.5&0.110 & 0.128 &0.081 & 0.093 & -0.02 & 0.1 & 0.03 \\
\hline
3 &0.079  &0.09  & 0.060& 0.069 & -0.015 & 0.07 & 0.02 \\
\hline
3.5& 0.0586 & 0.075 &0.047 & 0.053 &-0.01 & 0.06 & 0.015 \\
\hline
4 & 0.044&0.06  & 0.037 &0.042 &-0.009 & 0.04 & 0.01 \\
\hline
\end{tabular}
\caption{The values  of the $G_A$ form factors at high values of $Q^2$
\label{tab:2}}
\end{center}
\end{table}

\section{Summary}
In summary, in present work we calculated the axial-vector form factor of nucleon in the framework hard-wall AdS/QCD model. The hard-wall result of the axial-vector form factor obtained here is in a good agreement with the existing experimental data and with the results of different theoretical approaches.

{\bf Acknowledgements}

S. Mamedov and B.B. Sirvanli thanks T. Aliev for useful discussions. S. Mamedov thanks S. Siwach for reading manuscript and useful comments. This work has been done under the grant 2221 - Fellowships for Visiting Scientists and Scientists on Sabbatical Leave of TUBITAK organization of Turkey.


\begin{thebibliography}{99}
\bibitem{1}
  J.~M.~Maldacena,  Adv.\ Theor.\ Math.\ Phys.\  {\bf 2}, 231 (1998)
\bibitem{2}
  J.~M.~Maldacena,  Int.\ J.\ Theor.\ Phys.\  {\bf 38}, 1113 (1999)
  [arXiv:9711200[hep-th]]
\bibitem{3}
  S.~S.~Gubser, I.~R.~Klebanov and A.~M.~Polyakov, Phys.\ Lett.\ B {\bf 428}, 105 (1998)
  [arXiv:9802109[hep-th]].
\bibitem{4}
  E.~Witten,  Adv.\ Theor.\ Math.\ Phys.\  {\bf 2}, 253 (1998)
  [arXiv: 9802150[hep-th]].
\bibitem{5}  H. Boschi-Filho and N.R.F. Braga, JHEP {\bf 0305} (2003) 009
 [arXiv:0212207[hep-th]]
\bibitem{6}  H. Boschi-Filho and N.R.F. Braga, Eur.\ Phys.\ J. C {\bf 32} (2004) 529, [arXiv: 0209080[hep-th]]
\bibitem{7}
  J.~Erlich, E.~Katz, D.~T.~Son and M.~A.~Stephanov, Phys.\ Rev.\ Lett.\  {\bf 95}, 261602 (2005)  [arxiv:0501128[hep-ph]],
\bibitem{8}
  L.~Da Rold and A.~Pomarol,  Nucl.\ Phys.\ B {\bf 721}, 79 (2005) [hep-ph/0501218],
\bibitem{9}  L.~Da Rold and A.~Pomarol,  JHEP {\bf 0601}, 157 (2006)
  [hep-ph/0510268].
\bibitem{10} A.~Karch, E.~Katz, D.~T.~Son and M.~A.~Stephanov, Phys.\ Rev.\ D {\bf 74}, 015005 (2006) [arxiv:0602229[hep-ph]].
\bibitem{11} S.J. Brodsky, G. F. de Teramond, H.G. Dosch and J. Erlich, Light-Front Holographic QCD and Emerging Confinement  [arXiv:1407.8131 [hep-ph]]
\bibitem{12} Z. Abidin and C.Carlson, Phys. Rev. D {\bf 79}, 115003 (2009),[arXiv:0903.4818[hep-ph]]
\bibitem{13} H.R. Grigoryan and A.V. Radyushkin, Phys.\ Lett. B {\bf 650}, 421 (2007)[ arXiv:0703069 [hep-ph]]
\bibitem{14} N. Maru and M. Tachibana, Eur. Phys. J. C {\bf63}, 123 (2009) [arXiv:0904.3816[hep-ph]]
\bibitem{15} D.K. Hong, T. Inami and H.-U. Yee, \ Phys.Lett. B {\bf 646}, 165 (2007) [arXiv:0609270[hep-ph]]
\bibitem{16} T. Gutsche, V.E. Lyubovitskij, I. Schmidt and A. Vega, Phys. Rev. D {\bf 86}, 036007, (2012) [arXiv:1204.6612 [hep-ph]]
\bibitem{17} T. Gutsche, V.E. Lyubovitskij, I. Schmidt and A. Vega, Phys. Rev. D {\bf 87}, 016017, 2013, [arXiv:1212.6252[hep-ph]]
\bibitem{18} H.C. Ahn, D.K. Hong, C. Park and S. Siwach, Phys. Rev. D {\bf 80}, 054001 (2009) [arXiv:0904.3731[hep-ph]]
\bibitem{19} P. Colangelo, F. De Fazio, F. Giannuzzi, F. Jugeau and S. Nicotri, Phys.Rev. D{\bf78} (2008) 055009, [arXiv:0807.1054 [hep-ph]]
\bibitem{20} P. Colangelo, J.J. Sanz-Cillero, F. Zuo, JHEP {\bf 1211} (2012) 012, [arXiv:1207.5744 [hep-ph]]
\bibitem{21} N.Huseynova  and Sh. Mamedov, Int. J. Theor. Phys. {\bf 54} (2015) 3799
\bibitem{22} K.~Jo, B.~-H.~Lee, C.~Park and S.~-J.~Sin, JHEP {\bf 1006}, 022 (2010)
  [arXiv:0909.3914 [hep-ph]].
\bibitem{23}C.~Park, B.~-H.~Lee and S.~Shin, Phys.\ Rev.\ D {\bf 85}, 106005 (2012),[arXiv:1112.2177 [hep-th]]
\bibitem{24} B.-H. Lee, Sh. Mamedov, S. Nam and C. Park, JHEP {\bf 1308} (2013) 045 [arXiv:1305.7281[hep-th]]
\bibitem{25} B.-H. Lee, Sh. Mamedov and C. Park, Int. Jour. Mod. Phys. A {\bf 29} (2014) 1450170 [arXiv:1402.6061[hep-th]],
\bibitem{26} Sh. Mamedov,  Eur.Phys.J. C{\bf 76} (2016) 83, [arXiv:1504.05687 [hep-th]]
\bibitem{27} C.P. Herzog, Phys.Rev.Lett. {\bf 98} (2007) 091601,[arXiv:0608151 [hep-th]];
\bibitem{28}  P. Colangelo, F. Giannuzzi and S. Nicotri, Phys.Rev. D{\bf 80} (2009) 094019
[arXiv:0909.1534 [hep-ph]];
\bibitem{29} A.S. Miranda, C.A. Ballon Bayona, H. Boschi-Filho and N.R.F. Braga, JHEP {\bf 0911} (2009) 119, [arXiv:0909.1790 [hep-th]];
\bibitem{30} P. Colangelo, F. Giannuzzi, S. Nicotri and V. Tangorra, Eur.Phys.J. C{\bf 72} (2012) 2096
[arXiv:1112.4402 [hep-ph]];
\bibitem{31} P. Colangelo, F. Giannuzzi and S. Nicotri, JHEP {\bf 1205} (2012) 076, [arXiv:1201.1564 [hep-ph]];
\bibitem{32} L. Bellantuono, P. Colangelo, F. De Fazio and F. Giannuzzi, JHEP {\bf 1507} (2015) 053 [arXiv:1503.01977 [hep-ph]]
\bibitem{33} S. Sachan and S. Siwach,  Mod.Phys.Lett. A{\bf27} (2012) 1250163
[arXiv: 1109.5523 [hep-th]]
\bibitem{34} C. Park, Phys. Lett. B{\bf760} (2016) 79, [arXiv: 1603.04101[hep-th]]
\bibitem{35} V. Bernard, L. Elouadrhiri and U.-G. Meissner, J.Phys. G{\bf28} R1 (2002) [arXiv: 0107088 [hep-ph]]
 \bibitem{36} C. Alexandrou, M. Brinet, J.Carbonell, M. Constantinou, P. A. Harraud, P. Guichon, K. Jansen, T. Korzec and M. Papinutto,  	Phys. Rev. D{\bf83} 045010 (2011)
     [arXiv: 1012.0857 [hep-ph]]
\bibitem{37} G. Eichmann, H. Sanchis-Alepuz, R. Williams, R. Alkofer and C. S. Fischer, Baryons as relativistic three-quark bound states, Prog.Part.Nucl.Phys. {\bf91} (2016) 1, [arxiv:1606.09602 [hep-ph]]
\bibitem{38} I.V. Anikin, V.M. Braun and N. Offen, Phys. Rev. D {\bf94} 034011 (2016), [arXiv: 1607.01504 [hep-ph]]
\bibitem{39} A. S. Meyer, M. Betancourt, R. Gran and R.J. Hill, Phys.Rev. D{\bf93} (2016) 11, 113015 [arXiv:1603.03048 [hep-ph]]
\bibitem{40} K. Park et al. [CLAS Collaboration], Phys. Rev. C {\bf85}, 035208
(2012) [arXiv: 1201.0903 [nucl-ex]]
\bibitem{41} M.R. Schindler, T. Fuchs, J. Gegelia and S. Scherer, Phys.Rev. C{\bf 75} (2007) 025202 [arXiv: 0611083 [nucl-th]]
\bibitem{42} M. Henningson and K. Sfetsos, Phys. Lett. B{\bf 431}, 63
(1998) [arXiv:hep-th/9803251].
\bibitem{43} W. Mueck and K. S. Viswanathan, Phys. Rev. D{\bf 58}, 106006 (1998) [arXiv:hep-th/9805145].
\bibitem{44} R. Contino and A. Pomarol, JHEP{\bf 0411}, 058 (2004)
[arXiv:hep-th/0406257].
\bibitem{45} M. Henneaux "Boundary terms in the AdS / CFT correspondence for spinor fields" Proceedings of the International Workshop "Mathematical methods in modern theoretical physics" p. 161, Tbilisi, 1998; ULB-TH-99-03, [arXiv:9902137 [hep-th]]
\bibitem{46} D.K. Hong, H.-C. Kim, S. Siwach and H.-U. Yee, JHEP {\bf 0711} 036  (2007)
[arXiv:0709.0314 [hep-ph]]
\bibitem{47}  V. M. Braun, D. Yu. Ivanov, A. Lenz and A. Peters, Phys. Rev. D {\bf 75}, 014021 (2007).
\end{thebibliography}
\end{document}